\documentclass[numberedreferences]{kluwer}
\usepackage{graphicx}

\def\kms{km~s$^{-1}$}
\begin{document}
\begin{article}
\begin{opening}
\title{PSR B1849+00 probes the tiny-scale molecular gas?}            

\author{S. \surname{Stanimirovi\'{c}}\email{sstanimi@astro.berkeley.edu}}
\institute{Radio Astronomy Lab, UC Berkeley, 601 Campbell Hall, Berkeley, 
CA 94720}
\author{J. M. \surname{Weisberg}} 
\institute{Department of Physics and Astronomy, Carleton College, Northfield,
MN 55057}                               
\author{J. M. \surname{Dickey}}
\institute{Department of Astronomy, University of Minnesota, 116 Church St. SE,
Minneapolis, MN 55455}
\author{A. \surname{de la Fuente}}
\author{K. \surname{Devine}}
\author{A. \surname{Hedden}}
\institute{Department of Physics and Astronomy, Carleton College, Northfield,
MN 55057} 
\author{S. B. \surname{Anderson}}
\institute{Department of Astronomy, MS 18-34, California Institute of
Technology, Pasadena, CA 91125}


\runningtitle{PSR B1849+00 probes the tiny-scale molecular gas?}
\runningauthor{Stanimirovi\'{c} et al.}

\begin{abstract} 
In this paper we present and discuss the great difference in 
OH absorption spectra against PSR B1849+00 and SNR G33.6+0.1 
along the same line-of-sight. This finding is important as it 
clearly demonstrates that statistics of absorbing molecular gas
depends on the size of the background source. 
\end{abstract}


\end{opening}

\section{Introduction}

Studies of the absorption of signals from background continuum
sources by the intervening medium have been a very powerful way of
probing the properties of the interstellar medium (ISM). 
Pulsars are particularly suitable as background sources because of 
their pulsed radiation which allows us to investigate,
in both emission and absorption, {\it almost exactly the same
line-of-sight} (Weisberg et al. 1995; Koribalski et al. 1995). 
Another great advantage pulsars have is that their
continuum emission subtends over an extremely small solid angle, 
allowing us to probe needle-thin samples of the ISM.

Motivated by the pulsars' unique capabilities for studying the ISM, 
we measured the absorption spectra of several pulsars at the 
wavelength of the hydroxyl radical (OH), $\lambda = 18$ cm, using 
the Arecibo telescope. We detected OH absorption against one of our 
sources -- PSR B1849+00. The line-of-sight toward B1849+00 is 
particularly interesting as it passes right through the Galactic plane
($b=0^{\circ}$) and is very close (8 arcmin south) 
to a nearby supernova remnant (SNR) G33.6+0.1.

\section{Observations and Data Processing}

OH observations were undertaken with the Arecibo (305 m) radio 
telescope\footnote{The Arecibo Observatory is part of the National Astronomy
and Ionosphere Center, operated by Cornell University under a
cooperative agreement with the National Science Foundation.}.
The FWHM of the Arecibo telescope beam is approximately $2'.6 \times 3'.0$ 
at 1.6 GHz.
The Caltech Baseband Recorder was used as a fast-sampling backend, 
simultaneously covering both OH mainlines (at 1665 and 1667 MHz). 
Two types of spectra of astrophysical interest were formed during the
off-line data processing stage: the pulsar absorption spectrum, 
which depicts the pulsar signal alone (as absorbed by any intervening OH);
and the ``pulsar--off'' spectrum, which registers all emission 
and absorption lying in the telescope beam during the time that the 
pulsar signal is not present. The final spectra have velocity 
resolution of 0.9 \kms.
More information about our observations and data processing can be found in
Stanimirovic et al. (2003).

\section{Pulsar and SNR OH Absorption Spectra}

\begin{figure}
\centerline{\includegraphics[width=3.7in]{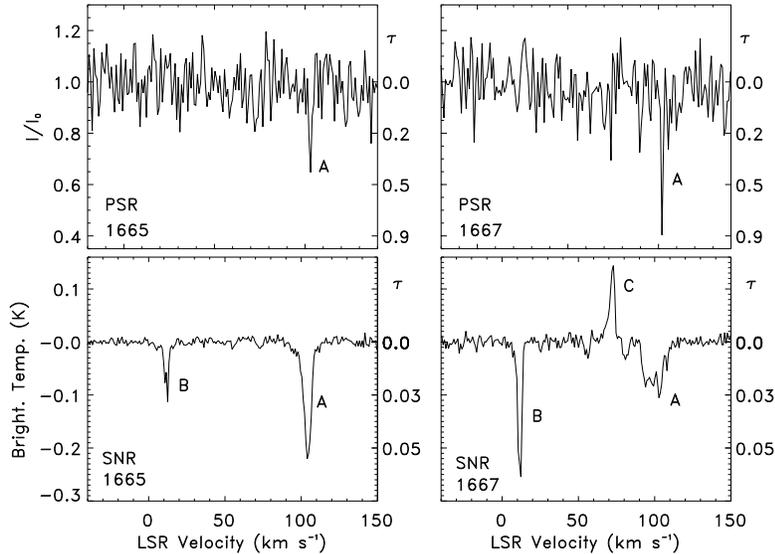}}
\vspace{0.7cm}
\caption{{\it Top two panels:} 
Pulsar absorption spectra toward B1849+00 produced against the pulsar
continuum emission alone at 1665 and 1667 MHz. 
{\it Bottom two panels:} Pulsar-off spectra toward B1849+00 produced
against the continuum emission from G33.6+0.1 at 1665 and 1667 MHz. In
addition to the absorption system at 102 \kms, an absorption system at 10
\kms~(`B') and an emission feature at 70 \kms~(`C') are seen.}
\end{figure}

Pulsar OH absorption spectra at 1665 and 1667 MHz are shown in Fig. 1 (top
two panels). At both frequencies, narrow absorption lines were 
detected at a velocity of about 102 \kms. 
The {\it only} previous OH absorption detected against a pulsar at 1667
MHz, to our knowledge, was by Slysh (1972).
Spectra in Fig. 1 depict the  pulsar signal {\it{alone}} as being 
absorbed  by intervening OH.  
The absorption system shown in Fig. 1, which we label as `A',
has higher optical depth than what is typically found towards
extragalactic sources (Dickey et al. 1981; Colgan et al. 1989). 
The ratio of the equivalent widths for the 1665 and 1667 MHz lines is
however very close to 5:9 which is expected for thermalized level populations.

The pulsar-off spectra are presented in the same figure (bottom two panels). 
As the SNR G33.6+0.1 is partially covered by the 
Arecibo beam absorption features in these spectra are effectively produced
against the continuum emissiom from G33.6+0.1. 
The absorption features at 102 \kms~differ greatly, 
in both peak intensity and linewidth, from
corresponding features seen in the pulsar absorption spectra at the 
same velocity.
In particular, feature `A' in the 1667 MHz line is almost 15 times
wider and 30 times shallower that its
corresponding feature in the PSR absorption spectrum.

\section{Possible Geometrical Explanations}

\begin{figure}
\centerline{\includegraphics[width=3.3in]{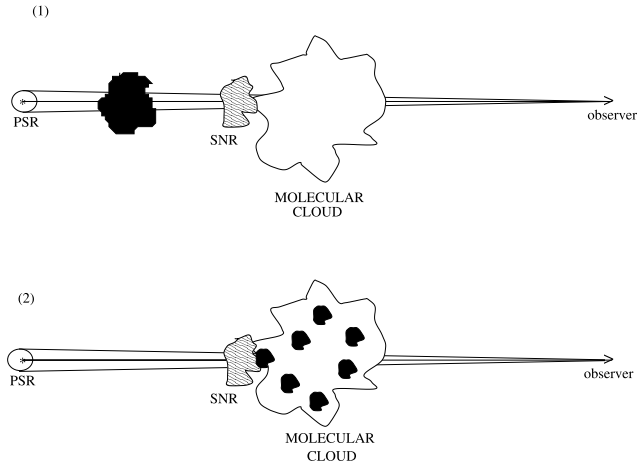}}
\caption{Two possible scenarios for the origin of OH absorption lines.
(1) An additional molecular cloud, shown in black, is located in front of
the PSR but behind the SNR. (2) All molecular gas is located in front of
both the SNR and the PSR. The PSR absorption is produced by a small
molecular clump (`cloudlet'), while the SNR absorption spectrum is produced
by an ensemble of small clumps (shown in black).}
\end{figure}

The pulsar absorption spectra are very deep and narrow, tracing
dense molecular gas with $N_{\rm H} \sim 10^{23}$ cm$^{-2}$, 
if the OH excitation
temperature $T_{\rm ex}=10$ K is assumed. 
However, the most striking observational result is that 
the pulsar absorption and pulsar-off spectra
appear to trace very different absorption features 
along {\it the same} line-of-sight and with the same central 
velocity of 102 \kms.
This result has to account for two additional constraints. 
First, a large molecular cloud was observed in $^{12}$CO(1-0) in 
the direction toward the SNR and the PSR by Green \& Dewdney (1992). 
Second, it was suggested that the two objects are
interacting with each other (Green 1989; Green \& Dewdney 1992).
Below we investigate two different geometrical scenarios that can explain 
the large difference in OH optical depths found against the PSR and the SNR.

(1) {\it An additional molecular cloud could be located in front of the PSR
yet behind the SNR.}
This would be the simplest explanation whereby OH absorption features in
the PSR and the pulsar-off spectra originate from two physically unrelated
molecular clouds (see Fig. 2, case 1). The sharp and deep absorption lines,
seen in the PSR absorption spectra, are produced
by an additional molecular cloud located in front of the PSR yet behind the
SNR. On the other hand,  broad OH absorption lines in the pulsar-off
spectrum are most likely due to the 
the interaction between the SNR with the molecular cloud.

This additional molecular cloud located behind the SNR could be of any size.
However, the large-scale $^{12}$CO(1-0) distribution presented in 
Green \& Dewdney (1992) does not show any obvious features that 
could be associated with this secondary cloud. 
In addition, we compared the hydrogen column densities derived
from OH and $^{12}$CO(1-0) in the PSR direction and found a good agreement.
This suggests that, most likely, all OH seen in
absorption and CO seen in emission coexist in
the same region making the existence of an additional molecular cloud along
the line of sight unlikely.

(2) {\it All molecular gas is in front of the SNR and the PSR.}
An alternative possibility is that the OH absorption features seen in the PSR
absorption and pulsar-off spectra originate from {\it the same} 
general molecular cloud located in front of both the PSR and the SNR.
This could happen in the case where the PSR absorption is produced by a 
small clump (`cloudlet'), while the shallower, broader absorption 
features against the SNR are caused by an ensemble of `cloudlets' of 
varying properties (Fig. 2, case 2).
More interestingly, the small `cloudlet' could represent a typical building 
block for the molecular cloud.

The solid angle subtended by the `cloudlet' intercepts
solid angles of both PSR and SNR continuum emission regions, however 
the pulsar-off spectrum does not appear to have a significant 
contribution from the `cloudlet'. 
This suggests that the `cloudlet' covers a very small fraction of the SNR and 
can be used to place an upper limit on its size. 
By assuming that the molecular cloud is at the distance of 7 kpc, we
estimate that the `cloudlet' radius must be $< 1$ pc while
its hydrogen volume density is $n>10^{5}$ cm$^{-3}$.

\section{Discussion}

As discussed in the previous section
PSR absorption spectra reveal existence of fine spatial structure 
in the absorbing
OH gas on scales $<1$ pc.  Also, all molecular gas seen in absorption is 
most likely located in front of both the SNR and the PSR. 
This is a clear demonstration that a pencil-sharp 
OH absorption sample against the PSR differs {\it dramatically} from a 
large-angle absorption sample against the SNR.  
The example of B1849+00 and G33.6+0.1 shows that
measured optical depths in OH depend heavily on the size of the 
background source.
This OH result is very different from  HI absorption findings 
(Dickey et al. 1979; Dickey et al. 1981; Payne al. 1982)
where absorption statistics was compared for a wide range of angular size
sources and no significant difference was found.
This led to the conclusion that the 
`cloudlet' model of the interstellar HI, whereby HI clouds are composed of
a large number of randomly distributed smaller clumps (or  `cloudlets'), 
is not prominent. However, the difference at HI and OH is not
totally unexpected: the solid-angle effect is expected to be
more pronounced for molecular gas where clumpiness is known to be significant.

We have investigated whether the PSR OH optical depth profiles 
could be building blocks for the molecular cloud by
modeling the SNR optical depth profiles with
an ensemble of PSR profiles (see Stanimirovic et al. 2003).
It was shown that the `cloudlet' model is not appropriate.
A more complex structure of the molecular cloud 
is required to explain the observed OH line profiles. 
In order to constrain better molecular cloud geometry
detailed OH observations of the whole SNR are crucial.
Another open question is 
whether the line-of-sight toward B1849+00 and G33.6+0.1
is unique or similar examples exist elsewhere in the ISM. We
would like to encourage further OH observations towards pulsars to constrain
how common this phenomenon may be.

\section{Conclusions}

We presented here the second ever  
detection of the OH absorption against a pulsar.
Absorption lines were detected against PSR B1849+00
in both OH mainlines. In addition we
detected OH absorption against a nearby SNR, G33.6+0.1.
The two sets of absorption profiles differ greatly but
most likely trace the same molecular cloud located
in front of both the SNR and the PSR. This surprising result
indirectly points to the existence of small scale ($<1$ pc)
structure in the absorbing OH gas. Also, it shows that
angular size of background sources can influence
greatly optical depth measurements in OH. This is opposite to
what was found for the HI absorbing gas.

\begin{acknowledgements}
This work was supported in part by NSF grants  AST-0097417 and AST-9981308.
\end{acknowledgements}

\end{article}
\end{document}